# RAY TRACING METHOD FOR STEREO IMAGE SYNTHESIS USING CUDA

Al-Oraiqat Anas M. [1*], Zori S. A. [2]

[*1]Taibah University, Department of Computer Sciences & Information
Kingdom of Saudi Arabia, p.o.Box 2898
Email: anas_oraiqat@hotmail.com
[2]Donetsk National Technical University, 83000, Donetsk, Artyoma 58, Ukraine
Email: sa.zori1968@gmail.com

**Abstract** - This paper presents a realization of the approach to spatial 3D stereo of visualization of 3D images with use parallel Graphics processing unit (GPU). The experiments of realization of synthesis of images of a 3D stage by a method of trace of beams on GPU with Compute Unified Device Architecture (CUDA) have shown that 60 % of the time is spent for the decision of a computing problem approximately, the major part of time (40 %) is spent for transfer of data between the central processing unit and GPU for computations and the organization process of visualization. The study of the influence of increase in the size of the GPU network at the speed of computations showed importance of the correct task of structure of formation of the parallel computer network and general mechanism of parallelization.

## 1. Introduction

Tasks of realistic visualization, generation of both static and dynamic realistic images of objects, and scenes of a surrounding situation by computer systems is a very active research domain that led to several works where it moves to a new qualitative level of volume visualization gradually. The modern computer systems of synthesis and visualization of a surrounding situation demand high quality effective application and a combination of methods of realistic 3D graphics as with the traditional mechanism of visualization, the most widespread and most available today is the stereo 3D visualization. Applications of volume visualization treat are: traditional computer graphics, any other areas of scientific research, practical activities that require improvements, and new quality level of display in results of computer simulation and modeling [1].

The above discussion gives rise to new updates and direction of applied research in the direction of creating effective architectures of software and hardware systems for solving realistic volumetric imaging. A huge research is conducted on the area of the task of realistic spatial visualization on devices of volumetric display; hence its result is intensively implemented in real world applications. Furthermore, the development of effective methods and means volume (space) 3D visualization is an active domain for researchers. However, there is no general and efficient approach of implementing neither quality surround display nor actual devices for their realization [1-5].

The organization of computer systems for realistic 3D spatial visualization provides a fundamentally new organization of the computational process, as compared to the standard 3D graphics pipeline. In the standard 3D graphics pipeline, the use of complex methods of synthesis and visualization (such as ray tracing, etc.) increases the realism of spatial 3D synthesis. In this regard, modern computer systems for image synthesis and visualization of the environment require a high-quality and effective methods and effective hardware support - graphics multiprocessors and multi-cores CPUs.

The purpose of this paper is to determine the efficiency 3D a stereo of visualization of 3D scenes with using ray tracing technique on the parallel graphic processing method with use of the Compute Unified Device Architecture technology (CUDA).

The rest of the paper is organized as follows. Section 2 gives the main features of volume visualization. Next, Section 3 introduces stereo visualization by method of tracing beams on parallel GPU. The experimental studies of the visualization system using GPU based ray tracing method and discusses the obtained results are presented in section 4. Finally Section 5 concludes the presented work.

## 2. Main features of volume visualization

Usually, the classical 3D graphics deals with virtual 3D space, the synthesized image, depending on the observer's position and visualized on a flat two-dimensional surface of the screen. At the same time synthesis of the image in fact consists in computation of a projection of a scene to the screen plane, and the image itself synthesized, as well as a way of its display (visualization), is flat, two-dimensional. Currently, there are two basic ways to display 3D information in bulk form 3D spatial visualization: 3D volumetric visualization and 3D stereoscopic visualization [1-5]. The main difference is that the spatial volume 3D visualization is not actually required to perform scene projection plane of the screen and execute a set of procedures of conventional 3D computer graphics, and the computation of the scene is in fact in the creation of the sampled 3D models of its 3D objects - visual 3D image objects in the scene, which is subsequently visualized on specialized devices with spatial and 3D volume display capabilities.

Stereoscopic 3D visualization essentially uses modified procedures of classical 3D computer graphics, which consists of computation of two projections of a scene (a stereo pair of images) to the display screen plane from two cameras, corresponding to eyes of the observer, double rendering of a scene, with further display of the received images on 3D displays. It should be noted that the existing 3D spatial visualization systems are costly and do not yet allow creating full quality physical objects described by the mathematical models created by methods of 3D graphics. Furthermore, one of the important reasons why devices on the basis of volume technologies of visualization are not widely used is lack of standardization of those technologies. The majority of volume 3D images spatially are now visualized by means of 3D methods based on stereoscopy visualization methods, which is easier in realization, and is a basic method of creating volume images for 3D devices - visualization at the present stage [1-5]. It is also necessary to note the following - for receiving correct and qualitative visual results today at creation of computer systems as spatial volume, and 3D visualization mainly use Ray tracing methods - "classical" Ray tracing and Ray casting in systems "classical" computer 3D schedules and 3D visualization, and "volume" method Volume Ray casting volume rendering for the organization in spatial systems voxel volume 3D visualization [6-7].

### 3. Stereo visualization by method of trace of beams on parallel GPU

Ray tracing - the method of computer graphics which allows creating photo realistic images of any 3D scenes, has been used successfully in computer graphics for a long time. Modern ray tracing algorithms are optimized versions of the basic algorithms, while using the reverse ray tracing algorithm (ray casting) because of the large computational load of the classical method [6] [8-10].There are many implementations of ray tracing optimization, the characteristics of some popular optimization ray tracing techniques are summarized in table 1 [9-10].

Table 1 - Comparative characteristic of modern methods of ray-tracing

| Method | Advantages | Shortcomings |
|---|---|---|
| Kd-trees | - Allows the use of binary search to find the primitive intersected by the ray.<br>- Simple and effective traverse algorithm.<br>- Good for the GPU.<br>- Require very little memory. | - Time-consuming construction, specially, search of splitting with the minimum SAH.<br>- Has a greater depth than the BVH. More steps in the construction. |
| Batch tracing | - Compatible other methods.<br>- Tracing group, which reduces the number of computations. | - A separate beam cannot be traced. |
| BVH-trees | - Possess a high speed and adaptability and construction.<br>- Fairly simple traverse for ray tracing, and for collision detection.<br>- Good for the GPU. | - Complexity of the data structures. |

However, even when using the modified and optimized trace methods, without using high-performance parallel computing systems the solution of a problem of synthesis in real time isn't possible [11-15]. Due to the huge volume of realistic visualization similar computations; rendering is divided into sub parallel streams. Therefore, using a multiprocessing system is an advantage to speed up the process visualization [14-15].

Both the direct and inverse trace of beams methods are well scaled on number of processors. As beams and photons can be traced almost independently from each other, that each node (processor) of parallel system can process a part of the image - for example, dividing the picture to N identical parts and charging to each processor to render the part [14-15]. This is shown in Figure 1.

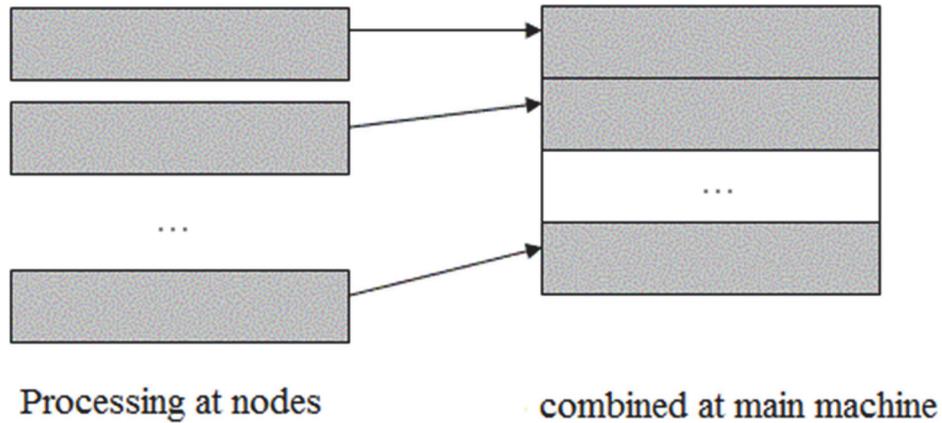

Figure 1 – Scheme of parallel tracing beams

    Modern direction for the implementation of resource-intensive computing is the use of high-performance capabilities of specialized parallel computing systems based on graphics display adapters. One such technology is the technology of CUDA on the graphics card NVIDIA. Thus, the basic procedure for the synthesis of a stereo image is a preparation (computation) stereoscopic (left image - for the left eye of the observer, and the right - for the right) [5] [14] [17]. Because computation of images can be performed absolutely independently for the left and right image, processes can be parallelized in better time and displayed on the parallel computer architectures. Due to the high popularity, availability and power the modern technology CUDA, that makes use of the capabilities of high-performance parallel graphics multiprocessor, computer synthesis of a stereo pair can be arranged as follows:

Parallel implementation of an independent synthesis "left channel" - "right channel" on multiprocessor (parallel level 1); Parallel "intracanal" implementation rendering using ray tracing on multiprocessor cores allocated for each channel (parallel level 2). Further, with the use of GPU post-processing process, stereo pair frames can also be carried out (frame conversions to display on the device of spatial visualization - stereo pairs building - for example, anaglyph/Anamorphic transformation. The implementation scheme of synthesis stereo is shown in Figure 2.

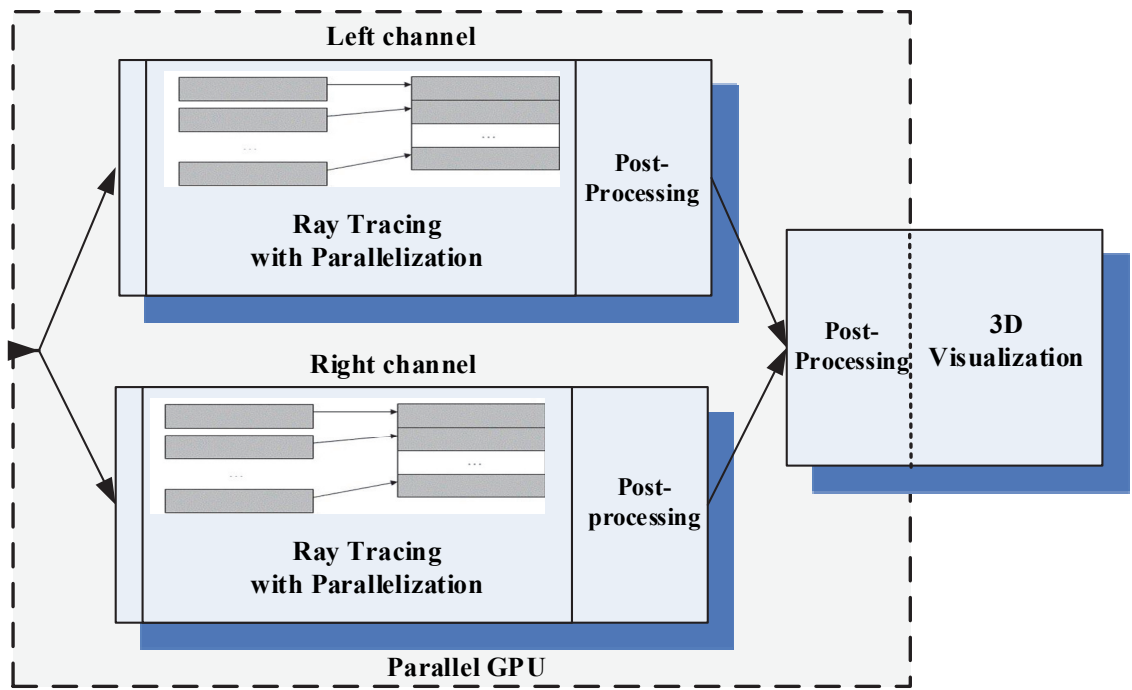

Figure 2 - The process of stereo synthesis using the GPU

## 4. Experimental study of the visualization system using GPU based ray tracing method.

Several experiments were conducted to investigate the dependence of the time characteristics of the synthesis on parameters that define complexity of the generated scene - quantities of sources of lighting, amount of objects making a scene, their complexity (quantity of the approximating sides), etc. and also parameters of the computing environment - the size of computing CUDA of a network and some other parameters. The studied characteristics of the system are: the time performance of computing part, time necessary for visualization of a scene, and general time of synthesis of a scene.

During experiments time volume which the system spends for each synthesis stage was also defined, for identification of losses of time for transfer of data from random access memory of the computer to memory of the multiprocessor and synchronization of work of parts of the system. Several experiments were conducted on the synthesis of 3D images on a video graphics card NVIDIA with a developed prototype software system [14-17]. The experimental data of image synthesis and one consisting of two objects and a light source are shown in figure 3 and figure 4 respectively.

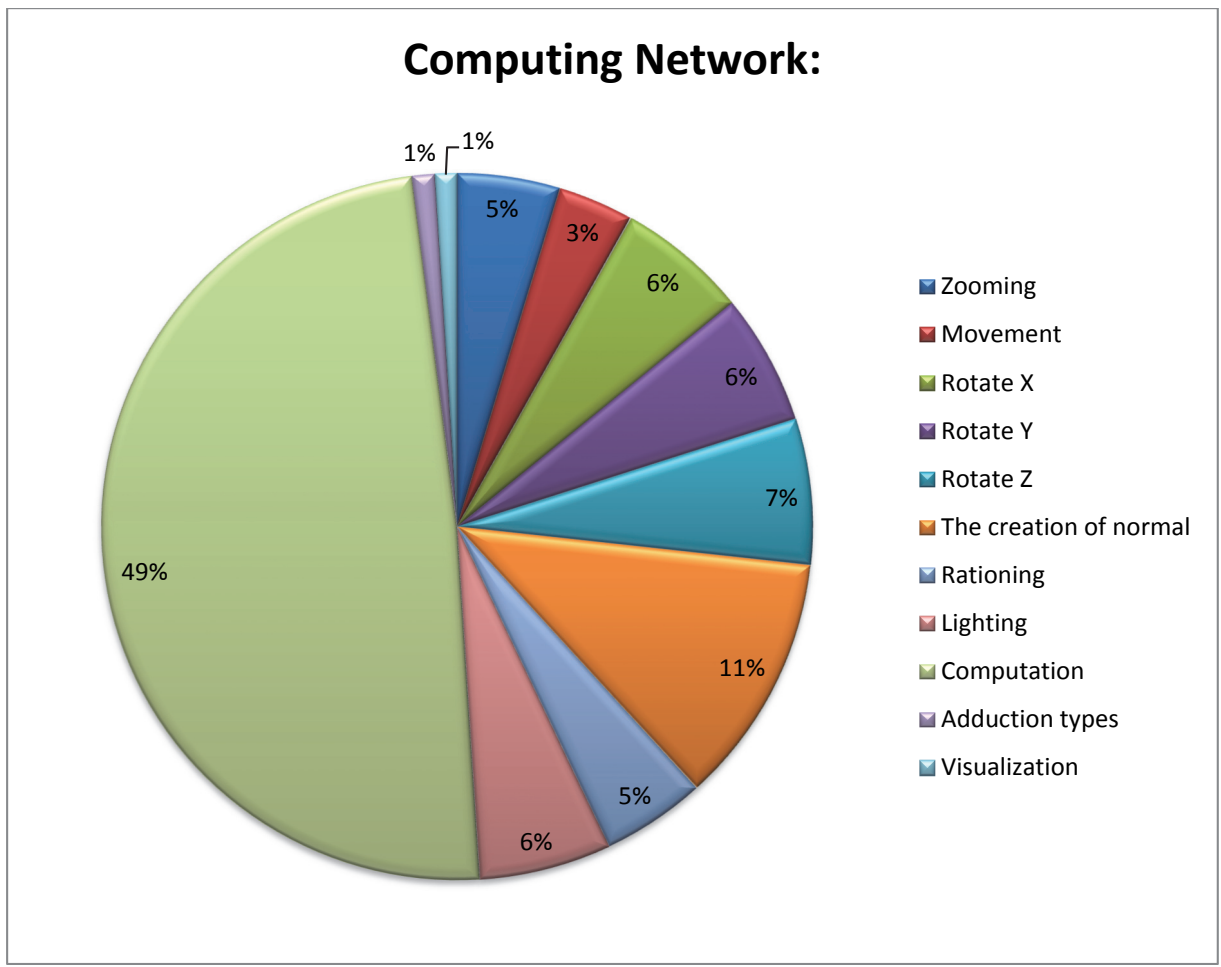

Figure 3 - Time functions of the synthesis of 3D scenes: 1 object, 1 light source

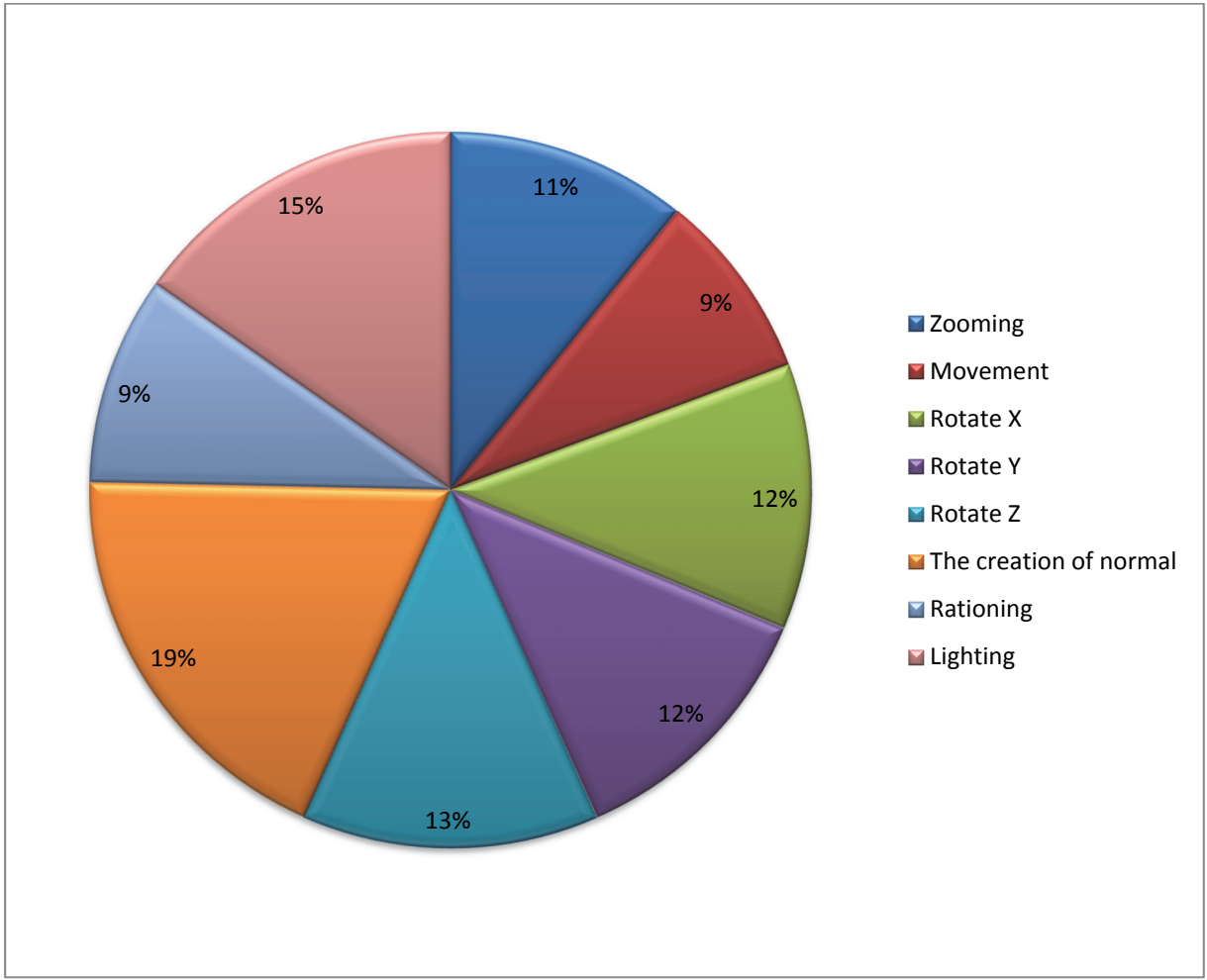

Figure 4 - Time functions of the synthesis of 3D scenes: 2 objects, 1 light source

Figure 5 shows a graph which illustrates the performance of the basic steps in the synthesis for various GPU network sizes (3 objects of the scene and 1 light source).

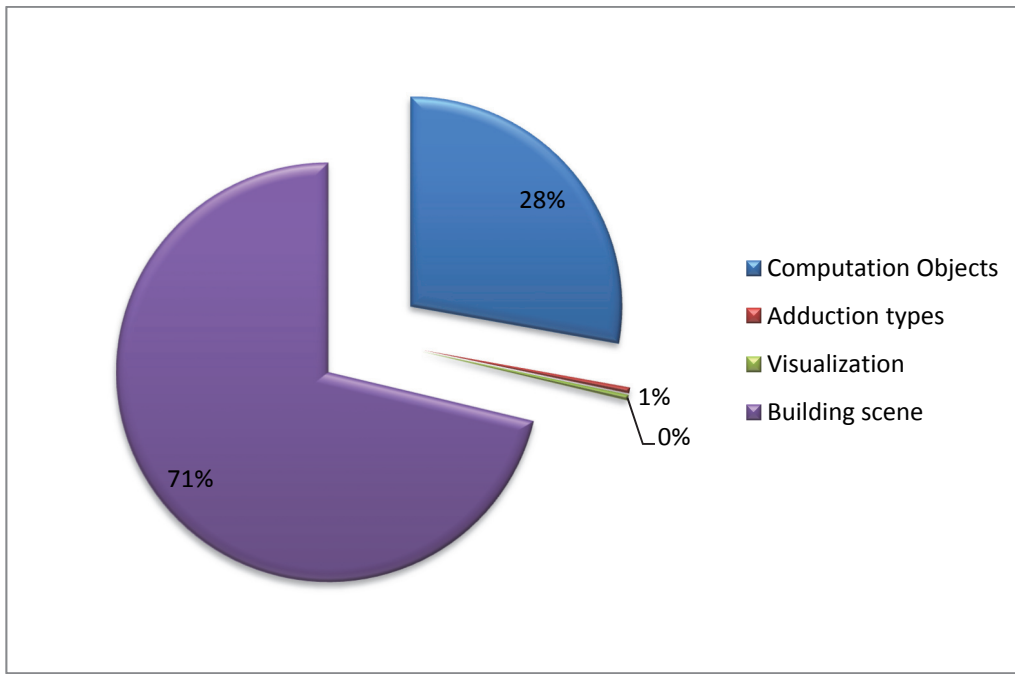

Figure 5 - Time functions of the synthesis of 3D scenes: 3 objects, 1 light source

Figure 6 shows the time required for scene computation consisting of five objects (two objects have 8 vertices and 12 polygons, two objects are composed of 12 nodes and 20 polygons, one object consists of 20 vertices and 36 polygons) and a single light source. The computations used GPU network sizes (1: 1).The results illustrate the influence of the size of the object on the computation speed.

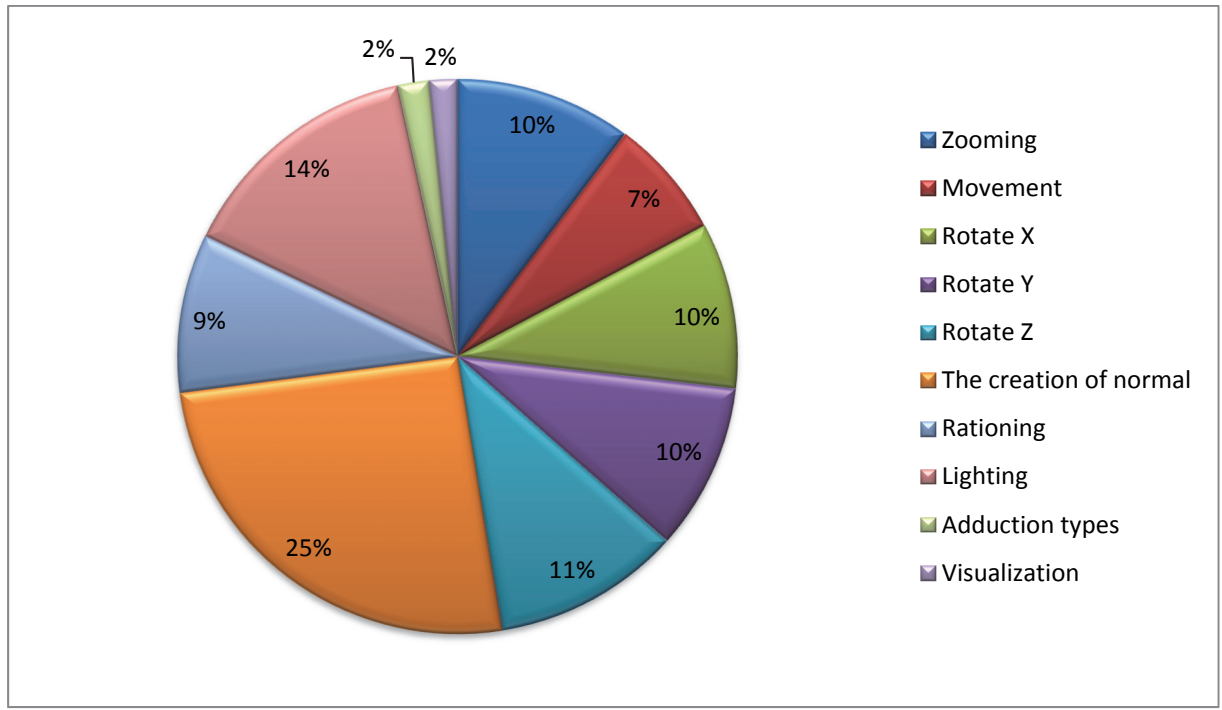

Figure 6 - Time functions of the synthesis of 3D scenes: 5 objects, 1 light source

Figure 7 shows the relative time spent on the basic procedures for the synthesis of a 3D image consisting of 6 objects with deferent complexity and one light source. It's clear that the computation part requires most of the time. In the "other" category falls the time taken to copy the data between the CPU and the GPU, and the time spent on the management of computing and visualization.

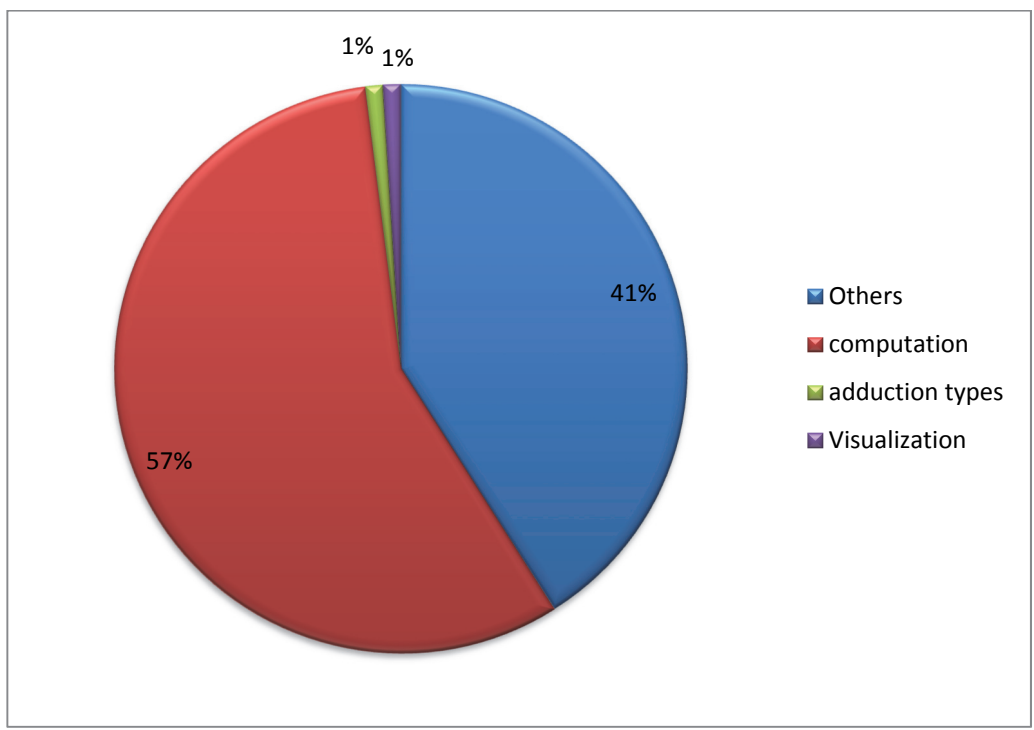

Figure 7 - Time spent on the basic procedures for the synthesis of images

Figure 8 illustrates the relationship between the times spent on the CUDA-computation on the network for a certain scene, depending on the chosen configuration of the network.

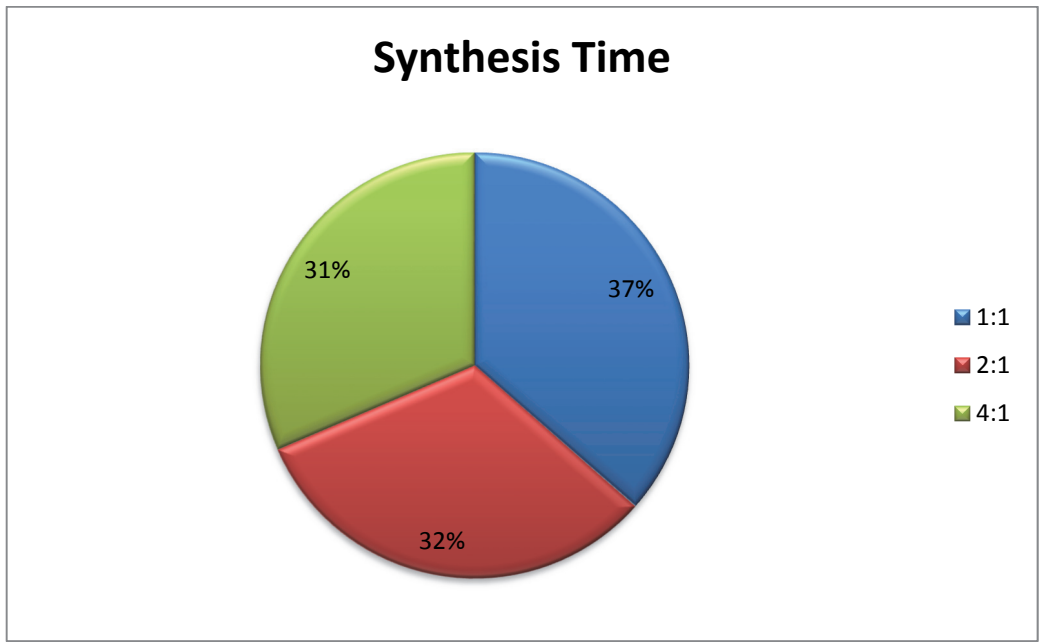

Figure 8 - The dependence of computation time on the network configuration

### 5. Conclusions

In this paper the problem of 3D spatial visualization of 3D scenes and the main approaches to its solution was considered. It is shown that the vast majority of volumetric 3D image space is visualized currently using the method of stereoscopic 3D visualization, as the easiest method to implement. It is shown that presently the main methods used in the synthesis of high-quality images, regardless of the chosen method of visualization, are the ray tracing methods. However, even with use of the modified and optimized trace methods, without the use of high-performance parallel computing solution to the problem of synthesis in real time is not possible. Therefore, it is required to visualize use of parallel computing systems.

The implementation of the approach of spatial 3D stereo visualization of 3D scenes with ray tracing using a parallel GPU is considered. The experiments showed that the developed prototype system solves the problem of image synthesis of simple 3D scene using ray tracing in few milliseconds. In this case:

-About 60 percent of the time is spent on the actual computation of the synthesis scene;

- A considerable part of task completion time (up to 40%) is spent on transfer of data between central and graphic processors at first for computations, and then for the organization of process of visualization;

- With the increase in computing complexity of a scene, time for its processing increases too; dependence has almost linear character;

- The solution of a test task on computing CUDA-of a network of the size (4:1) reduces time of computations for 20-25% in relation to a network of the size (2:1) and by 2.5 times in relation to a network (1:1);

- When organizing of process, the correct settings of the structure of formation of a parallel computing CUDA network and the general mechanism of parallelization are important.